# Intestinal gluconeogenesis and glucose transport according to body fuel availability in rats

Running title: Nutritional status and gut glucose metabolism

Caroline Habold[1], Charlotte Foltzer-Jourdainne[2], Yvon Le Maho[1],
Jean-Hervé Lignot[1] and Hugues Oudart[1]

[1]CNRS, CEPE, 23 rue Becquerel, F-67087 STRASBOURG cedex 2, France;
[2]INSERM, U. 381, 3 avenue Molière, F-67200 Strasbourg, France.

Correspondence to: Caroline Habold, CEPE-CNRS. 23 rue Becquerel. F-67087 STRASBOURG, France. E-mail address: caroline.habold@c-strasbourg.fr

## Summary

Intestinal hexose absorption and gluconeogenesis have been studied in relation to refeeding after two different fasting phases: a long period of protein sparing during which energy expenditure is derived from lipid oxidation (phase II), and a later phase characterized by a rise in plasma corticosterone triggering protein catabolism (phase III). Such a switch in body fuel uses, leading to changes in body reserves and gluconeogenic precursors could modulate the intestinal gluconeogenesis and glucose transport. The gene and protein levels, and the cellular localization of the sodium-glucose cotransporter SGLT1, and of GLUT5 and GLUT2 as well as that of the key gluconeogenic enzymes phosphoenolpyruvate carboxykinase (PEPCK) and glucose-6-phosphatase (Glc6Pase) were measured. PEPCK and Glc6Pase activities were also determined. In phase III fasted rats, SGLT1 was upregulated and intestinal glucose uptake rates were higher than in phase II fasted and fed rats. PEPCK and Glc6Pase mRNA, protein levels and activities also increased in phase III. GLUT5 and GLUT2 were downregulated throughout the fast, but increased after refeeding, with GLUT2 recruited to the apical membrane. The increase in SGLT1 expression during phase III may allow glucose absorption at low concentrations as soon as food is available. Furthermore, an increased epithelial permeability due to fasting may induce a paracellular movement of glucose. In the absence of intestinal GLUT2 during fasting, Glc6Pase could be involved in glucose release to the blood stream *via* membrane trafficking. Finally, refeeding triggered GLUT2 and GLUT5 synthesis and apical recruitment of GLUT2, to absorb larger amounts of hexoses.

## Introduction

Glucose is transferred across the brush-border membrane of enterocytes either by an energy-dependent sodium-dependent glucose cotransporter 1 (SGLT1) when its concentration does not exceed 30-50mM, or by diffusion at a higher concentration (Fullerton & Parsons, 1956; Debnam & Levin, 1975a; Lostao *et al.*, 1991). Fructose is transported by the facilitative transporter GLUT5 (Rand *et al.*, 1993). At the baso-lateral membrane of the cell, hexoses exit to the blood stream by facilitated diffusion *via* GLUT2 (Thorens *et al.*, 1990). GLUT2 can also be recruited to the brush-border membrane where it contributes to the absorption of glucose from the lumen (Kellett & Helliwell, 2000; Kellett, 2001).

Adult rats fed diets enriched in glucose show a specific upregulation of glucose transport activity in both the brush border membrane (Ferraris *et al.*, 1992) and the basolateral membrane (Cheeseman & Maenz, 1989). Also, during dietary restriction, the affinity of the carriers for sugars increases (Debnam & Levin, 1975b) and a 72h-fast causes an overall increase in glucose absorption in rats (Das *et al.*, 2001).

Fasting also induces an increase in gluconeogenesis in the small intestine which produces up to one third of endogenous glucose after 72 hours of fasting (Mithieux *et al.*, 2004). The main precursors of gluconeogenesis in the small intestine are glutamine and glycerol to a much lesser extent (Croset *et al.*, 2001). Two key gluconeogenic enzymes, the rate-limiting phosphoenolpyruvate-carboxykinase (PEPCK, EC 4.1.1.32) involved in glucose production from glutamine, and glucose-6-phosphatase (Glc6Pase, EC 3.1.3.9) which catalyzes the dephosphorylation of glucose-6-phosphate to glucose, are expressed in the small intestine and are upregulated in 48h fasted rats (Rajas *et al.*, 1999, 2000).

Studies on intestinal hexose absorption and on intestinal gluconeogenesis during food deprivation have hitherto been restricted to relatively short periods of fasting and have never been carried out according to the metabolic changes occurring through fasting. However, during fasting, the glucose sources and gluconeogenic precursors vary in relation with changes in body fuels utilization characterizing three distinct phases (Goodman *et al.*, 1980; Le Maho *et al.*, 1981): after a rapid period of adaptation marked by the depletion of glycogen reserves (phase I), lipid stores are progressively oxidized whereas body proteins are efficiently spared (phase II). The later fasting phase is characterized by both a strong increase in protein utilization as a substitute fuel for lipids (phase III) and a rise in plasma corticosterone level. Also, changing hormone levels through fasting may play a role in the regulation of gluconeogenic enzymes in the small intestine and on intestinal glucose transporters. The three phases have been first observed during fasting in penguins and rats (Le Maho *et al.*, 1981; Cherel & Le Maho, 1991a). Commonly, well-nourished humans do never reach a phase II fasting during interprandial periods. It is however noteworthy that fasting humans exhibit metabolic and hormonal changes characteristic of a phase II fast after a few days of fasting (Cahill *et al.*, 1966). In extremely malnourished humans like anorexic or cancer patients, a high urinary nitrogen loss and a low plasma fatty acid concentration, both characteristic of a phase III fasting, have been observed (Rigaud *et al.*, 2000). In obese patients submitted to total fasting, it has been shown that refeeding with a high carbohydrate diet was efficient and beneficial (Leiter & Marliss, 1983), but the intestinal cellular mechanisms implicated remain unknown.

The aim of this work, therefore, is to examine intestinal glucose absorption and production during distinct fasting phases and refeeding characterized by metabolic and hormonal changes (Goodman *et al.*, 1980; Le Maho *et al.*, 1981; Koubi *et al.*, 1991; Challet *et al.*, 1995).

Intestinal SGLT1, GLUT5 and GLUT2 gene and protein expressions were measured in phase II and phase III fasted rats. Since the energy load could first come from carbohydrate absorption after refeeding (Das *et al*., 2001), gene and protein expressions of these transporters were also studied in 2, 6 and 24h refed rats following either phase II, or phase III. To further examine glucose absorption, intestinal glucose uptakes were measured *in vivo* in normally-fed, phase II and phase III fasting rats. Finally, intestinal gluconeogenesis was studied during phase II, during which the glycerol released from lipid oxidation may be the main gluconeogenic precursor, and during phase III, when the delivery of amino acids to the gluconeogenic tissues increased markedly. In that way, PEPCK and Glc6Pase gene expressions, protein levels and activities were examined during the fasting phases and after refeeding.

## Methods

*Animals*

Male Wistar rats weighing 300g were obtained from Iffa-Credo (Lyon, France). Animals were housed individually in leucite cages with a wire mesh floor to prevent coprophagia, and maintained at 23°C with a 12-hour light period. They were fed a standard diet (A03 pellets from UAR, Epinay-sur-Orge, France) and had free access to water throughout the experiments. They were weighed every day between 9.00 and 10.00h *a.m*. Our experimental protocol followed the CNRS guide for care and use of laboratory animals.

*Experimental procedures*

After one week acclimatization, rats were killed as control animals (Ctrl), whereas the other rats were food deprived. The fasting phases (II, III) were determined by calculating the specific daily rate of body mass loss dM/Mdt (g/kg/day) for each animal (dM represents the loss of body mass during $dt=t_1-t_0$ and M is the rat body mass at $t_0$). The phase II fasting period lasting between one and six days (for 300g rats), a first group of rats was killed in the fourth day in phase II (P2r0, n=15). Three other groups were refed during 2h (P2r2, n=5), 6h (P2r6, n=5), or 24h (P2r24, n=5) following phase II, and then killed. Four additional groups continued fasting until the second day of phase III, reaching on average eight days of fasting. Such a fast is still reversible since animals in phase III can be successfully refed at this time, just as well after a spontaneous fast in wild animals as after an experimental fast (Le Maho *et al.*, 1976, 1981; Cherel & Le Maho, 1991a). In our study, one group was killed in phase III without refeeding (P3r0, n=15), whereas the three others were killed after refeeding for 2h (P3r2, n=5), 6h (P3r6, n=5), or 24h (P3r24, n=5) following phase III.

The animals were killed between 9.00 and 10.00 a.m. The jejunum was removed, weighed, and cut into segments. These segments were then treated separately depending on the analysis considered.

*Plasma parameters and intestinal enzyme activities*

Blood samples were collected immediately after sacrifice to measure plasma concentrations of urea, corticosterone, insulin and glucose in all experimental groups. Plasma urea was determined with an Urea Nitrogen Kit (Sigma Diagnostics, St Louis, USA) and plasma concentrations of corticosterone and insulin with Enzyme Immunoassay Kits [Assay Designs (Ann Arbor, USA) and Eurobio (Les Ulis, France), respectively]. Glycemia was measured using the glucose oxydase-peroxydase technique.

Glc6Pase activity was determined in the small intestinal mucosa as described by Rajas *et al.* (1999), and the Jomain-Baum and Schramm (1978) protocol was followed to measure PEPCK activity.

*Northern blot analysis*

Total RNA from duodenal and jejunal mucosa was isolated by the method of Chomczynski and Sacchi (1987). Of total RNA, 5µg was electrophoresed per lane in an agarose gel. After electrophoresis, RNA was transferred to a nylon membrane (Roche Diagnostics, Mannheim, Germany) by vacuum blotting and then fixed on the membrane by UV light. Blots were probed with specific digoxigenin-end-labeled (5') antisense oligonucleotide probes (Eurogentec, Seraing, Belgium) using the method of Trayhurn *et al.* (1995). The following probes were used: 5'-TGCCAGTCCCCCTGTGATGGTGTAAAGGGCGG-3' for SGLT1 (GenBank D16101), 5’-GGACTGGGCCCCACGGCGTGTCCTATGACGTA-3’ for GLUT5 (GenBank L05195), 5’-CCGCCCCGCCTTCTCCACAAGCAGCACAGAGA-3’ for GLUT2 (GenBank J03145), 5’-GGGTCAGCTCGGGGTTGCAGGCCCAGTTGTTG-3’ for PEPCK (GenBank XM_342593), 5’-CGGGACAGACAGACGTTCAGCTGCACAGCCCA-3’ for Glc6Pase (GenBank U07993). Slot-blots were stripped and re-probed with an 18S rRNA probe to correct for variations in RNA loading or blotting. The blots were analyzed by densitometry using Scion Image Software.

*Western blot analysis*

Total duodenal and jejunal proteins were isolated from the mucosa after centrifugation (90min, 30000g) in 10vol. (wt./vol.) ice-cold sample buffer [10mM Tris-HCl, 10% SDS, 15mg/mL DTT, 1% protease inhibitor cocktail (Sigma) pH 7.4], whereas the plasma membrane fractions were isolated by discontinuous sucrose density gradient centrifugation as described previously (McCartney & Cramb, 1993). Protein concentrations were determined by the Bicinchoninic Acid method. Western blotting was conducted using standard techniques (Hames, 1996). Proteins (25µg per lane) were separated by SDS-PAGE using 7% poly-acrylamide gels and electroblotted onto PVDF membranes before immunodetection processing. The membranes were incubated with the primary antibody [rabbit anti-rat SGLT1, Glut5, Glut2 (Chemicon, Temecula, USA), sheep anti-rat PEPCK (generously provided by Dr. D. K. Granner), rabbit anti-rat Glc6Pase (generously provided by Dr. G. Mithieux)]. Control blots were also run simultaneously using equivalent dilutions of either pre-immune serum or immune serum pre-incubated with the peptide antigen. Membranes were finally incubated with an alkaline phosphatase-conjugated secondary antibody and the bound antibodies were visualized by incubating the blots in BCIP-NBT (Chemicon). The level of immunoreactivity was then measured as peak intensity using an image capture and analysis system (Scion Image Analysis). Results were expressed as relative densitometric units, normalized to the values of Ponceau-stained blots to account for any differences in protein loading among lanes.

*Immunohistochemistry*

Formalin-fixed intestinal samples embedded in paraffin were cut 6-µm-thick and collected on poly-L-lysine-coated slides. Sections were cleared of paraffin, rehydrated and pre-incubated with a blocking solution containing normal goat serum. Sections were then incubated with the primary antibody (rabbit anti-rat SGLT1, Glut5, Glut2 and Glc6Pase) and finally, with Alexa Fluor-labelled goat anti-rabbit IgG (Molecular Probes, Eugene, USA). As controls, we used primary antibodies pre-absorbed with the respective peptides and preimmune serum. Sections were examined with a fluorescent microscope (Zeiss Axioplan 2).

*Glucose uptake measurements*

Glucose uptake rate into the small intestine was determined by using the *in vivo* perfused intestinal segments technique. Normally-fed and phase II and phase III fasted rats were anesthetized prior to surgery using IP sodium pentobarbital (60mg/kg body mass) and placed on a heated (37°C) surgical table. After performing a laparotomy, the small intestine was isolated and the luminal contents removed by gently flushing with saline solution at 37°C. An intestinal loop was cannulated and a recirculating perfusion was started at a flow rate of 2mL/min with 10mL saline solution at 37°C containing 5mM glucose. The animals were killed after an absorptive period of 20min. The small intestine was excised and the mucosa was scraped free of the underlying tissue and weighed. Glucose concentration was estimated in the luminal content using a glucose oxydase-peroxydase kit (Roche Diagnostics). The absorption rate was calculated from the difference between the total amount of glucose injected initially and that recovered after the end of experiment. Blood glucose concentration was also measured before and after luminal perfusion.

*Statistical analysis*

Data are presented +/- SEM. Statistical comparisons of experimental data were performed by one-way and two-way analysis of variance (ANOVA) and Tukey post-hoc test by using the software Sigmastat (Jandel). The level of statistical significance was set at $P<0.05$.

## Results

*Body mass loss*

The calculation of the specific daily rate of body mass loss dM/Mdt (g/kg/day) permitted the determination of the three fasting phases and a daily monitoring of the physiological status for each animal through fasting (figure 1). With this monitoring, all the animals survive the prolonged starvation procedure and can be successfully refed. The first fasting phase (phase I) lasted only a few hours and was characterized by a rapid decrease in dM/Mdt. The specific daily body mass loss then reached a steady rate (approximately 55g/kg/day) representing phase II and finally, strongly increased which was characteristic of phase III.

*Plasma parameters*

Plasma corticosterone was 375- and 29-fold higher in phase III fasted (P3r0) than in controls and P2r0 rats, respectively (table I). It diminished after only 2h refeeding.

Uremia did not vary between control and P2r0 rats (table I). In phase III fasting, plasma urea concentration was 3.75- and 3.5-fold higher than in control and P2r0 rats, respectively. After refeeding following a phase III, urea concentration progressively decreased compared to the P3r0 value.

Plasma insulin concentration dropped in phase II and phase III fasted rats (table I). It increased after 2h refeeding following phase II and after 6h refeeding following phase III but remained lower than in the control group even after 24h refeeding.

Glycemia decreased by 2.5-fold in phase II and by 4.5-fold in phase III fasted rats compared to controls (table I). While plasma glucose concentration rapidly increased after only 2h refeeding following phase II, it was still significantly lower in animals refed for 24h following a phase III than in control animals.

*SGLT1 gene and protein expressions*

The level of SGLT1 gene (figure 2A) and protein (figure 2B) expressions did not significantly vary between control, phase II fasted and refed rats following phase II. In phase III, SGLT1 gene expression increased 1.6- and 2.1-fold compared to the control and P2r0 values, respectively. Also, the amount of SGLT1 protein rapidly rose in phase III and was more than 3-fold higher than in control and phase II fasted rats. The high level of SGLT1 gene expression was maintained in refed animals following phase III, whereas its protein amount was lowered to control values. Changes in SGLT1 protein expression during fasting and after refeeding could also be observed through immunohistochemical labelling. As shown in figure 2C, SGLT1 was highly expressed in the brush border membrane of phase III fasted rats.

*GLUT5 gene and protein expressions*

While GLUT5 gene expression (figure 3A) did not vary, neither in phase II, nor in phase III fasted rats compared to controls, its protein amount (figure 3B) was significantly decreased by one half in these animals. The amounts of GLUT5 mRNA and protein were higher in rats refed following a phase III than following a phase II fast. GLUT5 immunolabelling (figure 3C) decreased during fasting, whereas it increased in the apical membrane of refed animals following phase III.

*GLUT2 gene and protein expressions*

GLUT2 gene expression (figure 4A) was decreased by 3.1- and 2.2-fold in phase II and phase III fasted rats, respectively, compared to controls, and its protein (figure 4B) was no more detectable in phase II and phase III fasted rats. After refeeding, the level of GLUT2 mRNA rapidly increased and the protein was already detected after 2h refeeding following either phase II, or phase III. The highest increase in GLUT2 mRNA and protein after refeeding could be observed in the P3r2 rats, where the values were no more different than the control ones. In control rats, GLUT2 was mainly localized in the enterocytes baso-lateral membranes (figure 4C). In refed animals following phase III, an intense staining of GLUT2 could be observed in the apical membrane.

*In vivo glucose uptake measurements*

Glucose absorption rate was expressed per milligram tissue and not per centimetre length, since intestinal villi and small intestine diameter were very reduced during fasting. Glucose was more absorbed in the small intestine of phase II and phase III fasted rats than of normally-fed rats (table II). Expressed per milligram of proteins, the results were virtually the same. Compared to normally-fed animals, glucose was 1.3- and 2.1- more absorbed in phase II and phase III fasted rats, respectively. The glycemia rose in glucose perfused fasted rats but remained still lower than in controls (table II).

*PEPCK gene expression, protein level and activity*

PEPCK mRNA (figure 5A) and protein levels (figure 5B) and PEPCK activity (figure 5C) did not vary between control, P2r0 and refed rats following phase II. However, a phase III fast induced a 4-, 13- and 30-fold increase in PEPCK mRNA, protein level and activity, respectively. These values remained still high after 2h refeeding following phase III. In these animals, the amount of PEPCK protein was even 2.5- and 32-fold higher compared to phase III fasted and control rats, respectively. This increase in PEPCK protein content in the P3r2 group was unexpected in regard to the changes in the levels of gene expression and activity and in regard to the values of the other experimental groups. PEPCK protein expression then decreased in rats refed 6h following phase III and was lower than in controls after 24h refeeding.

*Glc6Pase gene expression, protein level and activity*

Glc6Pase gene expression (figure 6A) did not vary between control and phase II fasted rats. The phase III fast induced a 10-, 2.3- and 2.1-fold increase in Glc6Pase mRNA, protein level (figure 6B) and activity (figure 6D), respectively. In refed animals following phase III, Glc6Pase mRNA level and activity decreased to control values, while no changes occurred after phase II. In refed animals following both phases, protein abundance decreased to levels lower than in controls. In control rats, Glc6Pase appeared to be in the epithelial cells along the whole villus axis and in all the intestinal villi (figure 6C). A cytoplasmic staining restricted to the apical part of the enterocytes immediately under the microvilli could be observed. In phase III fasted rats, the localization was similar but the staining appeared more intense.

## Discussion

To our knowledge, it is the first study to investigate intestinal glucose absorption and gluconeogenesis according to body fuel availability. It appeared that during protein catabolism (phase III fast), the active way of glucose absorption is induced by an increase in SGLT1 in the brush border membrane, whereas the facilitated component of intestinal glucose transport is triggered at refeeding by an increase in GLUT2 translocated to the apical membrane. The study also shows an increase in intestinal gluconeogenesis during the phase III fasting period, during which the gluconeogenic precursors may be amino acids coming from protein catabolism, compared to the phase II fasting period, *i.e.* when glycerol may be the main gluconeogenic precursor.

*Glucose transport*

While a phase of lipid oxidation (phase II) did not induce any changes in SGLT1 gene and protein expressions, their levels were upregulated by a phase of protein catabolism (phase III). These results suggest that the metabolic status of phase III fasted rats, rather than the duration of fasting, enhanced SGLT1 gene expression and protein synthesis. Such an increase in active glucose transporter level may allow intestinal glucose absorption immediately after refeeding even at low concentrations. As illustrated by the *in vivo* 5mM glucose uptake measurements, glucose is massively absorbed in the intestine of phase III fasted rats. It could involve SGLT1 and can be related to the high level of corticosterone, as reported in the normal (Batt & Peters, 1976; Batt & Scott, 1982; Iannoli *et al.*, 1998) and in the inflamed small intestine (Sundaram *et al.*, 1999). Noteworthy, in the inflamed intestine glucocorticoids are able to restore the levels of SGLT1 (Sundaram *et al.*, 1999). However, SGLT1 may not be solely involved in glucose absorption during fasting, since the 5mM glucose solution was also more absorbed during phase II fasting compared to normally-fed rats, without any increase in SGLT1 gene and protein expressions. An increase in intestinal permeability which allows paracellular movement of macromolecules has been observed during fasting and malnutrition (Worthington *et al.*, 1974; Welsh *et al.*, 1998; Boza *et al.*, 1999) and could therefore also permit glucose absorption during phase II and phase III fasting. An increase in glucose absorption during fasting has also been previously observed in dogs (Galassetti *et al.*, 1999). Refeeding following the phase III fast induced a decrease in SGLT1 protein amount which paralleled the decrease in plasma corticosterone concentration. In contrast, the level of SGLT1 mRNA remained unchanged after refeeding suggesting a post-transcriptional mechanism of regulation induced by refeeding. A post-transcriptional regulation of SGLT1 characterized by a higher stability of the mRNA induced by cAMP has already been showed in an in vitro renal model (Peng & Lever, 1995).

The amount of the facilitative hexose transporter GLUT5 protein was lowered during fasting and increased during refeeding. This is in accordance with previous studies showing that food ingestion involves *de novo* synthesis of GLUT5 mRNA and protein (Ferraris, 2001). However, the levels of mRNA did not parallel the level of protein except for the refeeding period following the phase III fast, suggesting that regulation of GLUT5 is at least partly post-transcriptional. It has been shown previously that the GLUT5 mRNA stability is increased by fructose, in part *via* an increase of the cAMP pathway (Gouyon *et al.*, 2003). In our study however, the decrease in GLUT5 protein content in fasting animals could be, at least partly, a non specific decrease, since during fasting the whole body protein synthesis is reduced (Cherel *et al.*, 1991b).

The protein and gene expressions of the facilitative glucose transporter GLUT2 were downregulated during phase II and phase III fasting and upregulated by refeeding. These changes may be linked to plasma corticosterone, shown to inhibit this transporter in stressed rats (Shepherd *et al.*, 2004). In refed rats, GLUT2 was recruited from the baso-lateral enterocytes membrane to the apical brush border membrane. Recruitment of GLUT2 to the apical membrane has been reported in previous studies and is partly controlled by the SGLT1-dependent activation of a protein kinase C (Kellett & Helliwell, 2000; Helliwell *et al.*, 2000a, 2000b). GLUT2 is characterized by a high Km and provides a passive component of glucose absorption from the intestinal lumen detectable at high concentrations (Kellett, 2001).

*Gluconeogenesis*

During the short phase I fasting period, glycogen stores are completely exhausted (Goodman *et al.*, 1980; Le Maho *et al.*, 1981). Glucose is then produced by gluconeogenesis from various precursors coming essentially from adipose tissue lipolysis during phase II (glycerol) and from protein catabolism during phase III. PEPCK and Glc6Pase mRNA levels and enzymatic activities have been shown to increase in the small intestine in 48h fasted rats and then to reach a plateau after 72h fasting (Mithieux *et al.*, 2004; Rajas *et al.*, 1999, 2000). In our study, PEPCK and Glc6Pase gene expressions, protein levels and activities did not vary during phase II fasting (5 days fasting) compared to normally-fed animals. Plasma levels of ketone bodies, which have been shown to inhibit specific CoA-dependent enzymes involved in gluconeogenesis in the liver (Balasse *et al.*, 1967; Kean & Pogson, 1979; Shaw & Wolfe, 1984; Fery *et al.*, 1996) are elevated during this fasting phase (Goodman *et al.*, 1980; Belkhou *et al.*, 1991). In phase II fasted rats however, plasma insulin (known to inhibit gluconeogenesis) dropped, whereas plasma corticosterone (known to stimulate gluconeogenesis) slightly increased. As a whole, these effects would lead to a lack of significant change in intestinal gluconeogenesis after a 5-day fast (phase II). During a phase III fasting period, the strong increase in gluconeogenesis may be a direct consequence of the elevated plasma corticosterone level and the decrease in ketone bodies (Goodman *et al*., 1980; Belkhou *et al*., 1991) in relation to the switch in the gluconeogenic precursors from glycerol to amino acids. Glucocorticoids are known to stimulate gluconeogenesis by induction of PEPCK (Friedman *et al*., 1993) and Glc6Pase (Ashmore *et al*., 1956; Nordlie *et al*., 1965; Voice *et al*., 1997) gene transcription in the liver. Also, corticosterone induced protein catabolism and thereby, the release of gluconeogenic amino acids. During phase III fasting, there is a concomitant increase in circulating amino acids and in glucagonaemia (Cherel *et al*., 1988), which could stimulate the uptake and conversion of plasma amino acids into glucose by the small intestine. Refeeding rapidly induced normalization of gluconeogenesis, despite a low glycemia. This might be explained by the rapid decrease in the level of plasma corticosterone and by the increase in the level of plasma insulin which is known to inhibit the gene expression of the gluconeogenic enzymes in the liver (Barthel & Schmoll, 2003). A decrease in Glc6Pase activity to base-line values has been shown previously in the liver (Newgard *et al*., 1984) and in the small intestine (Rajas *et al*., 1999) in rats refed after short fasting periods. Normalization of PEPCK activity is also rapidly achieved in the jejunum in these animals (Rajas *et al*., 2000).

Finally, the observation of Glc6Pase in the apical part of the enterocytes suggests that this enzyme could also be involved in the transepithelial transport of glucose, as shown previously (Stumpel *et al*., 2001; Santer *et al*., 2003). According to these studies, glucose absorbed through SGLT1 is phosphorylated into glucose-6 phosphate before entering the endoplasmic reticulum where it is hydrolyzed by Glc6Pase to glucose and phosphate. Glucose then re-enters the cytosol and diffuses out of the enterocytes through GLUT2 or is released into the blood stream by a membrane traffic pathway. This latter mechanism could permit glucose secretion into the blood when GLUT2 is absent like in phase II and phase III fasted rats.

In light of all this data, it appears that energy depletion in relation with body fuel uses and gluconeogenic precursors availability increases the ability of the intestine to absorb sugar from its lumen and stimulates intestinal gluconeogenesis. A phase II fast, associated with a high availability of glycerol as gluconeogenic precursor leads to an impaired ability of the small intestine to absorb glucose and to no change in intestinal gluconeogenesis. A phase III fast however, associated with a high availability of amino acids used as gluconeogenic precursors, leads to an increase in the ability of active glucose absorption. Glucose can then, be immediately absorbed at low concentrations at refeeding. At the same time, intestinal gluconeogenesis is increased. In the absence of GLUT2, Glc6Pase could also play a role in glucose transport through the cell and thereby, in its secretion into the blood stream. Finally, refeeding induces facilitative transport, so that large amounts of fructose and glucose can be transported from the intestinal lumen to the blood stream.

**Acknowledgments:**

We thank J.N. Freund, C. Domon-Dell and G. Mithieux for helpful discussion, and C. Arbiol for technical help. We are also grateful for G. Mithieux and D.K. Granner for providing us the Glc6Pase and PEPCK antibodies.

**Table I**

Plasma corticosterone, urea, insulin and glucose concentrations in control, fasted and refed rats

| | *Corticosterone* $10^{-7}$*g.L*$^{-1}$ | *Urea g.L*$^{-1}$ | *Insulin* $10^{-6}$*g.L*$^{-1}$ | *Glucose mM* |
|---|---|---|---|---|
| Ctrl | 0.83 ± 0.61[a] | 0.189 ± 0.007[a] | 3.20 ± 0.46[a] | 11.42 ± 0.73[a] |
| P2r0 | 10.77 ± 3.12[ab] | 0.203 ± 0.011[a] | 0.20 ± 0.06[b] | 4.57 ± 0.87[b] |
| P2r2 | 0.51 ± 0.29[a] | 0.319 ± 0.025[ab] | 1.20 ± 0.41[bc] | 9.88 ± 0.72[ac] |
| P2r6 | 1.60 ± 0.72[a] | 0.282 ± 0.014[ab] | 1.31 ± 0.46[bc] | 8.58 ± 0.54[cd] |
| P2r24 | 6.07 ± 2.43[ab] | 0.176 ± 0.015[a] | 1.84 ± 0.31[c] | 9.43 ± 0.27[ad] |
| P3r0 | 311.01 ± 62.69[b] | 0.708 ± 0.068[b] | 0.34 ± 0.11[b] | 2.57 ± 0.31[b] |
| P3r2 | 7.08 ± 1.75[ab] | 0.598 ± 0.067[b] | 0.39 ± 0.13[b] | 3.59 ± 0.27[b] |
| P3r6 | 14.67 ± 5.23[ab] | 0.448 ± 0.009[ab] | 1.33 ± 0.29[bc] | 4.22 ± 0.39[b] |
| P3r24 | 6.78 ± 3.84[ab] | 0.243 ± 0.017[ab] | 1.51 ± 0.26[bc] | 7.51 ± 0.68[d] |

Results are means ± SEM, n=6 rats per group. Within the same column, values with different letters are significantly different ($P<0.05$).

**Table II:**

*In vivo* glucose uptake measurements and glycemia after 5mM intestinal glucose perfusion in control, phase II and phase III fasted rats

| | Ctrl | P2r0 | P3r0 |
|---|---|---|---|
| Glucose absorption (5mM perfusate) ($10^{-4}$mmol glucose/mg mucosa/min) | 3.03 ± 0.23[a] | 3.91 ± 0.51[b] | 6.29 ± 0.56[c] |
| Glycemia mM (not perfused) | 11.42 ± 0.73[a] | 4.57 ± 0.87[c] | 2.57 ± 0.31[c] |
| Glycemia mM (5mM perfusate) | 14.33 ± 2.46[b] | 9.24 ± 0.97[d] | 6.27 ± 1.21[d] |

Results are means ± SEM, n=5 rats per group. Values with different letters are significantly different (P<0.05). Glucose absorption was compared according to the metabolic status (Ctrl, P2r0, P3r0). Glycemia was compared according to the metabolic status (Ctrl, P2r0, P3r0) and to the perfusion condition (none, 5mM perfusate).

**Figure 1:**

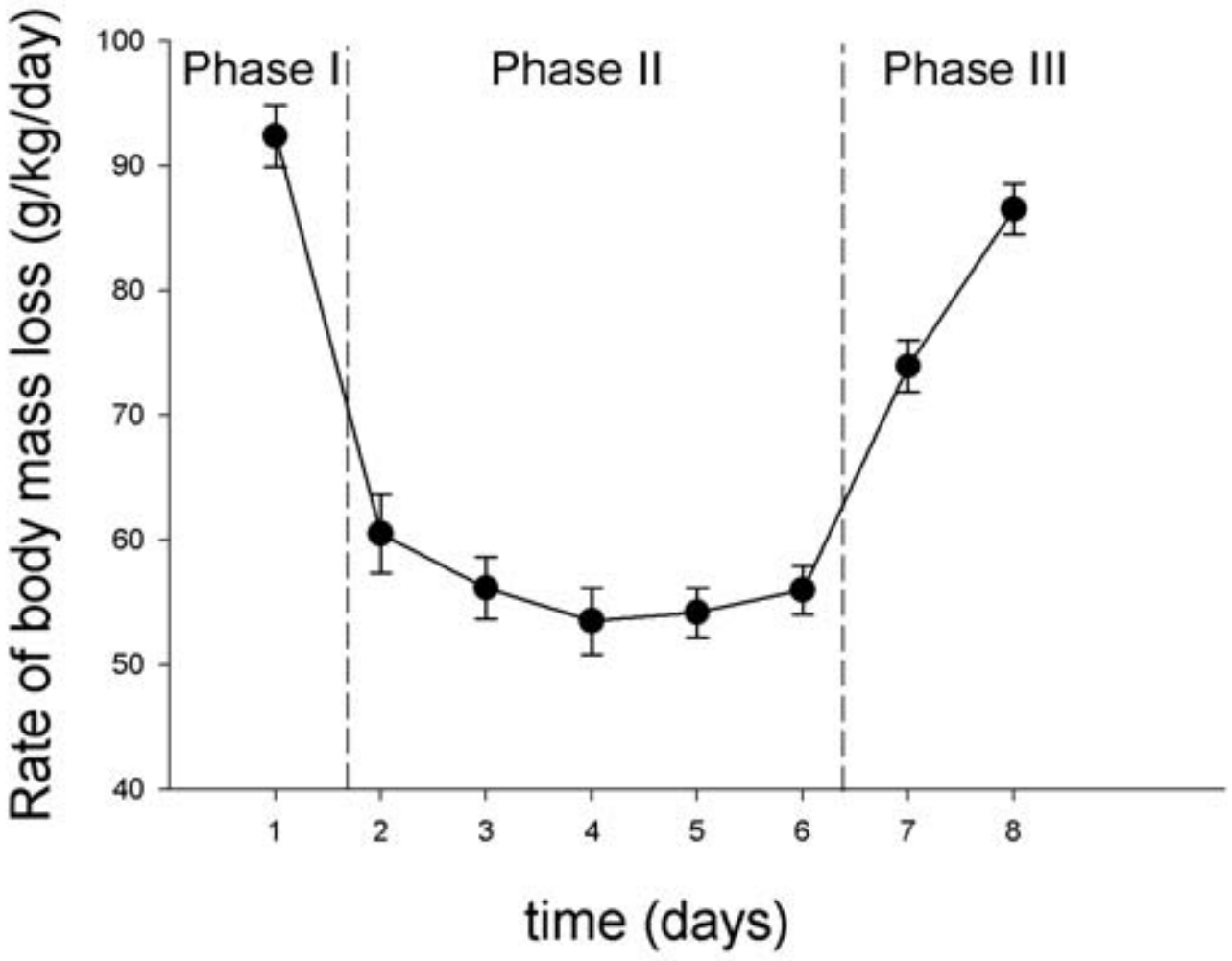

**Figure 2:**

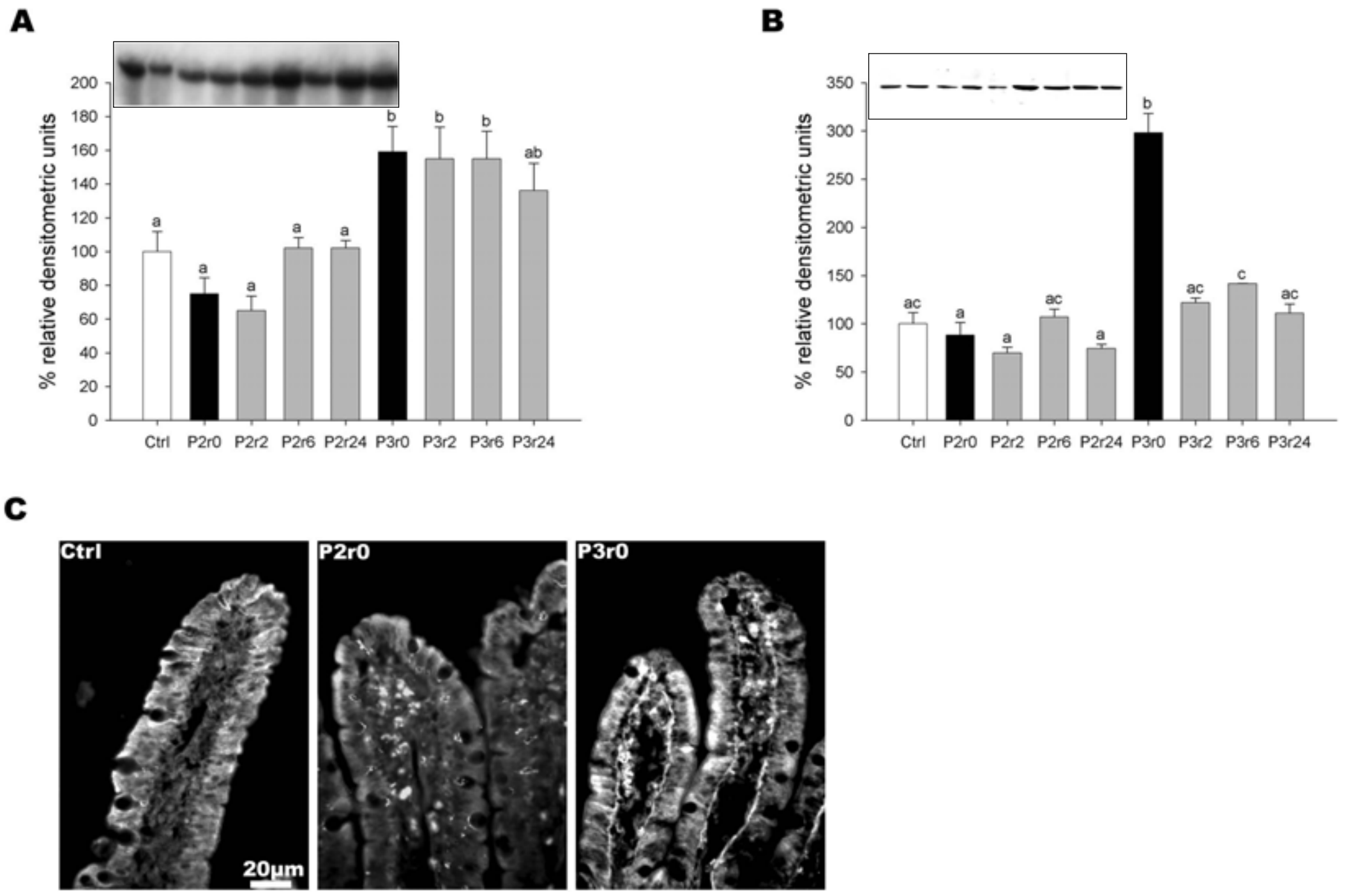

**Figure 3:**

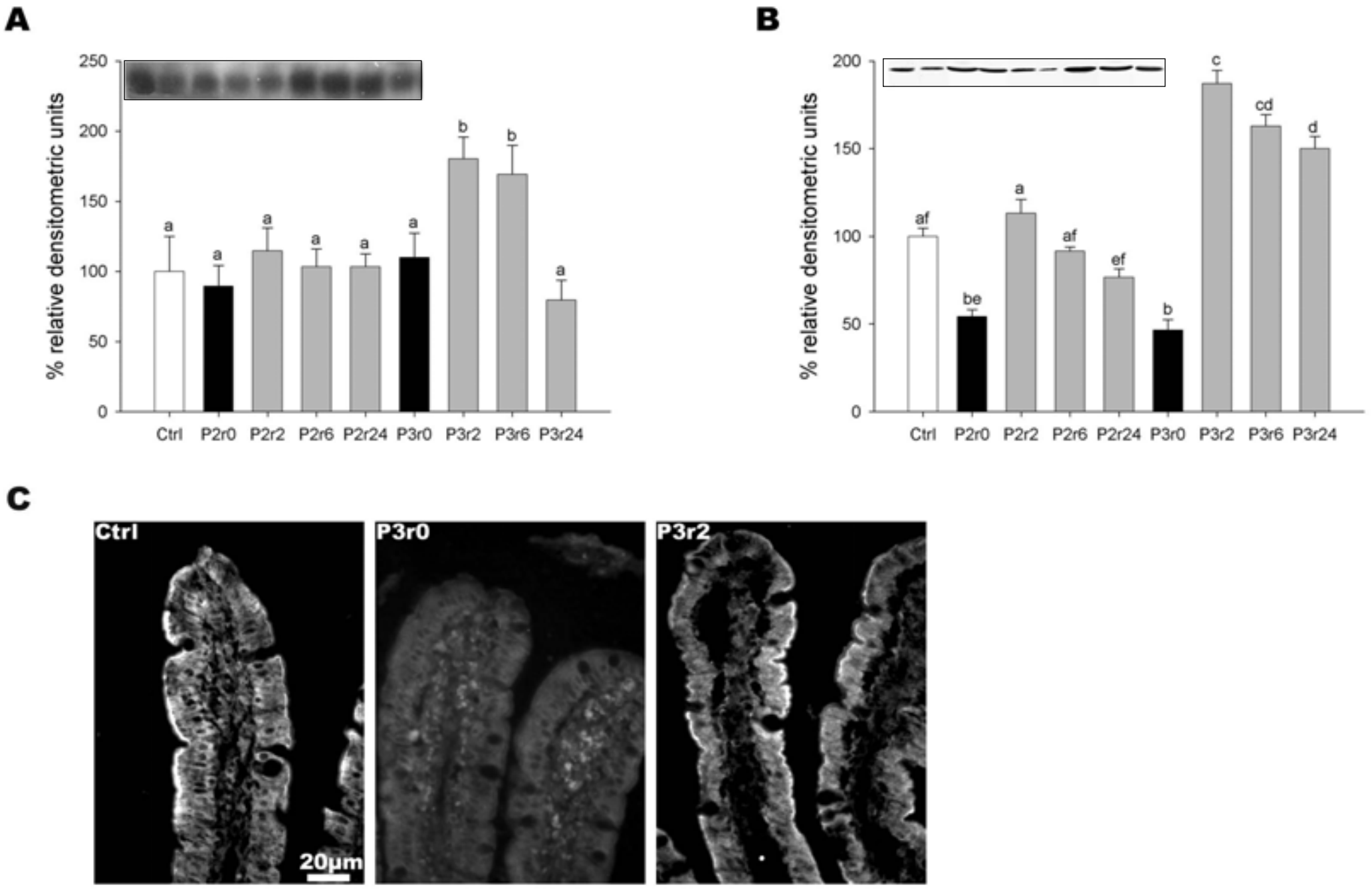

**Figure 4:**

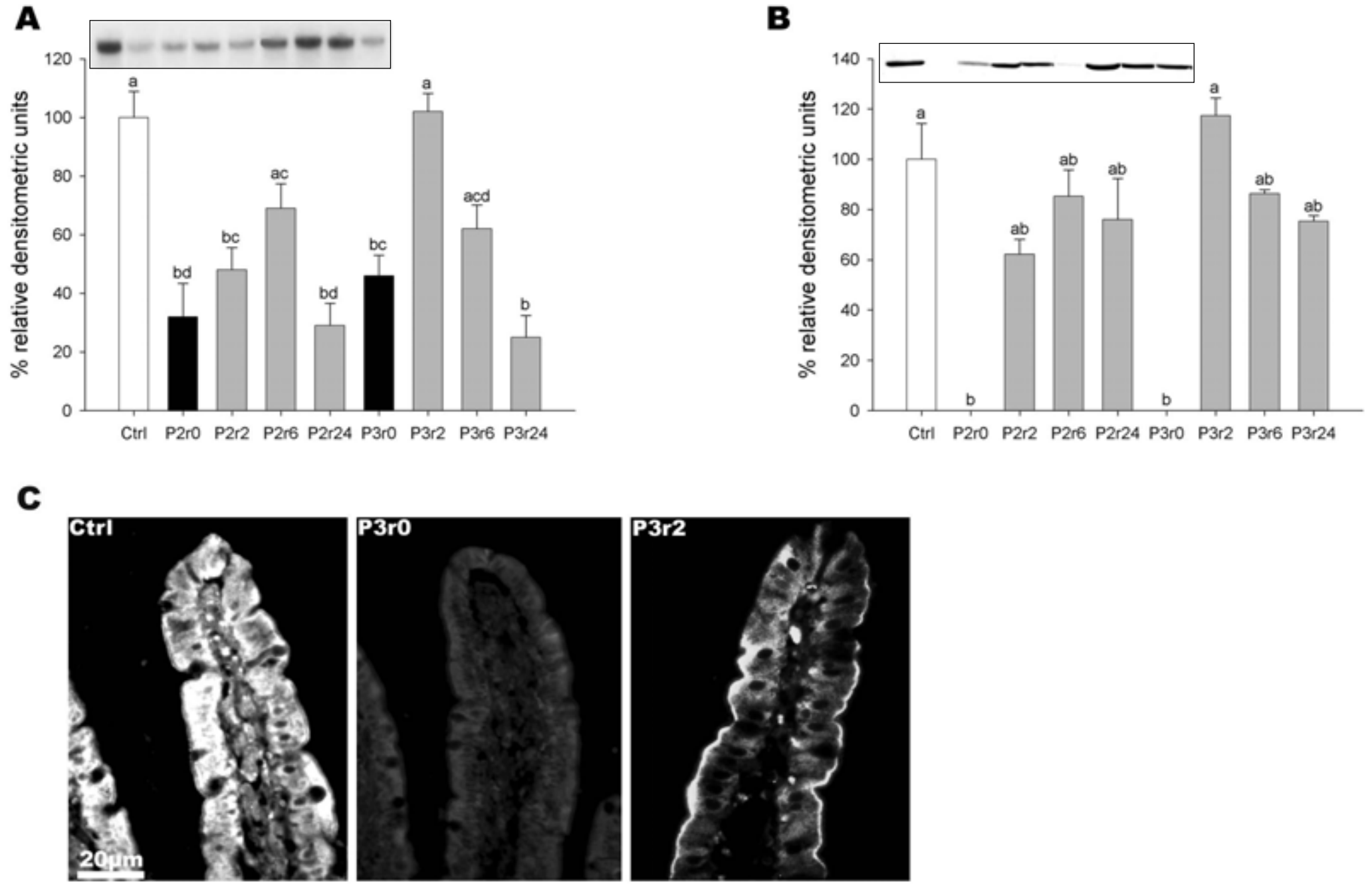

**Figure 5:**

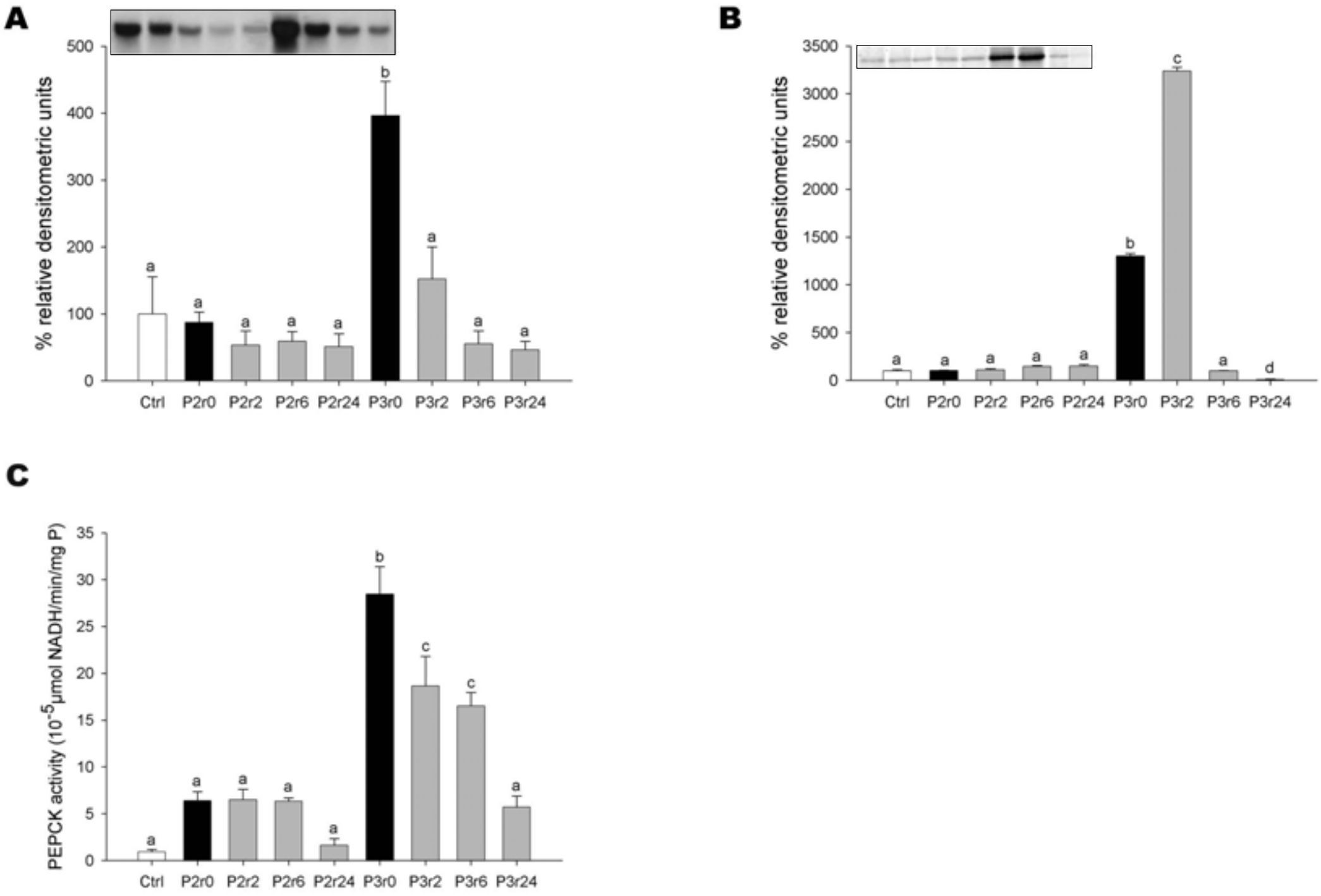

**Figure 6:**

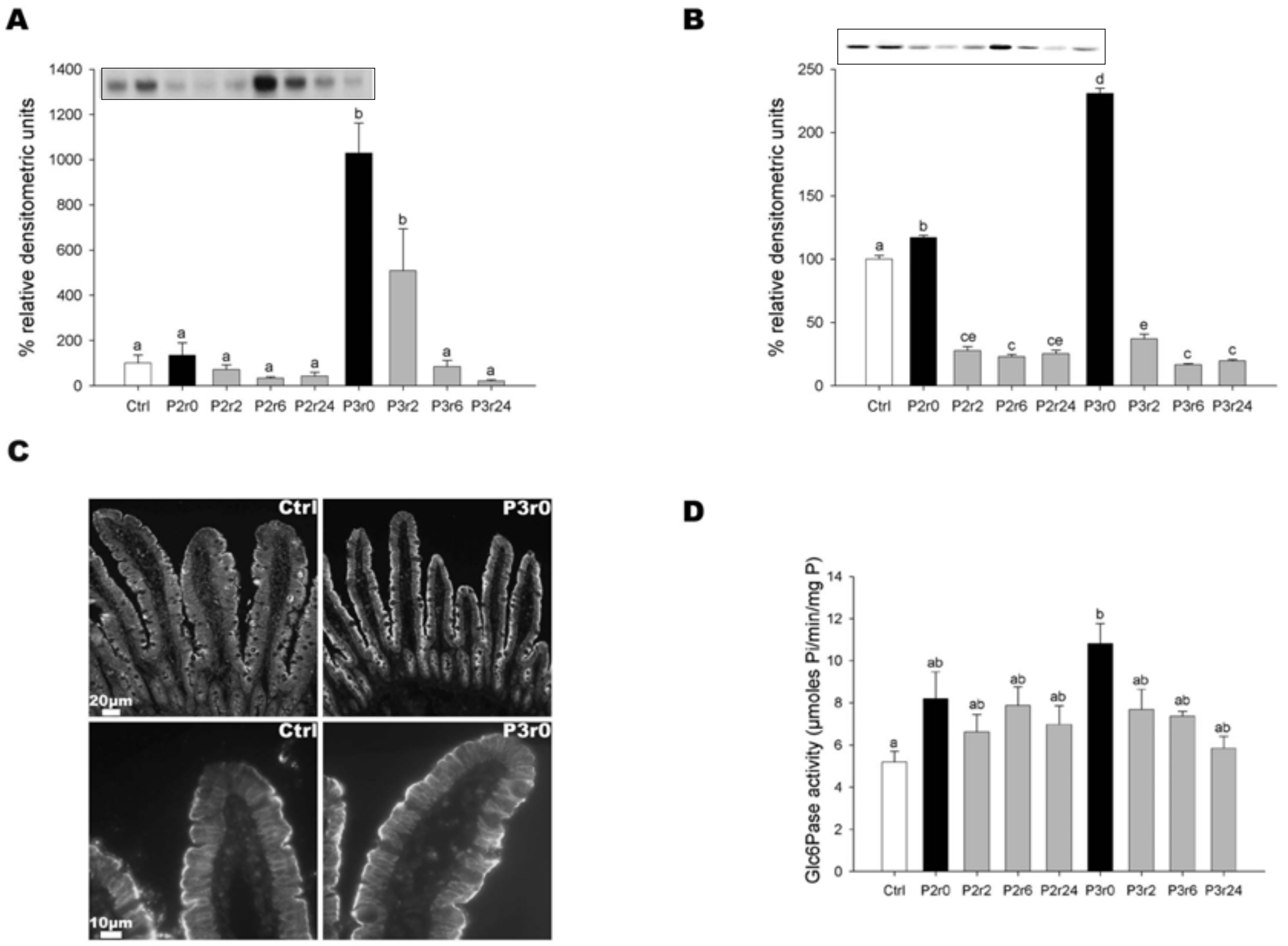

## Figure legends

Figure 1: Rate of body mass loss (dM/Mdt, g/kg/day) in fasted rats. Mean ± SEM (n=5).

Figure 2: SGLT1 gene (A) and protein (B) expressions in Ctrl, P2r0 and P3r0 rats and after refeeding (P2r2, P2r6, P2r24, P3r2, P3r6, P3r24). The upper panels show representative northern (A) and western (B) blots and the lower panels show the densitometric analysis (Ctrl to P3r24 from the left to the right). (C) shows SGLT1 immunolocalization in control, phase II and phase III fasted rats. The immunofluorescence in the *lamina propria* is non-specific. Mean +/- SEM (n=5 per group).

Figure 3: GLUT5 gene (A) and protein (B) expressions in Ctrl, P2r0 and P3r0 rats and after refeeding (P2r2, P2r6, P2r24, P3r2, P3r6, P3r24) and GLUT5 immunolocalization (C) in control, phase III fasted and 2h refed rats following phase III. Mean +/- SEM (n=5 per group).

Figure 4: GLUT2 gene (A) and protein (B) expressions in Ctrl, P2r0 and P3r0 rats and after refeeding (P2r2, P2r6, P2r24, P3r2, P3r6, P3r24) and GLUT2 immunolocalization (C) in control, phase III fasted and 2h refed rats following phase III. Mean +/- SEM (n=5 per group).

Figure 5: PEPCK gene expression (A), protein level (B) and activity (C) in Ctrl, P2r0 and P3r0 rats and after refeeding (P2r2, P2r6, P2r24, P3r2, P3r6, P3r24). Mean +/- SEM (n=5 per group).

Figure 6: Glc6Pase gene expression (A), protein level (B) and activity (D) in Ctrl, P2r0 and P3r0 rats and after refeeding (P2r2, P2r6, P2r24, P3r2, P3r6, P3r24). (C) shows Glc6Pase immunolocalization in control and phase III fasted rats. Mean +/- SEM (n=5 per group).